\documentclass{article}
\usepackage[utf8]{inputenc}
\usepackage[english]{babel}
\usepackage{siunitx}
\usepackage{lscape}
\usepackage{indentfirst}
\usepackage{lscape}
\usepackage{subcaption}
\usepackage{multirow}
\usepackage{multicol}
\usepackage{amsmath}
\usepackage{cleveref}
\usepackage{pdfpages}
\usepackage{authblk}
\usepackage{multicol}
\usepackage{gensymb}
\usepackage{wrapfig}
\usepackage{needspace}
\newenvironment{Figure}
  {\par\medskip\noindent\minipage{\linewidth}}
  {\endminipage\par\medskip}
\usepackage[font=footnotesize,labelfont=bf]{caption}

\usepackage{geometry}
\usepackage{graphicx}
\graphicspath{ {./images/} }

\usepackage[
backend=biber,
style=numeric,
sorting=none,
]{biblatex}
\usepackage{titlesec}

\titlespacing*{\maketitle}
{0pt}{1.5ex}{1ex}
\titlespacing*{\section}
{0pt}{1.5ex}{1ex}
\titlespacing*{\subsection}
{0pt}{1.5ex}{1ex}

\font\myfont=cmr16
\title{\vspace{0cm}\myfont{High-resolution laser system for the S$^3$-Low Energy Branch}}

\author[1]{\footnotesize{Jekabs Romans}}
\author[2]{\footnotesize{Anjali Ajayakumar}}
\author[3]{\footnotesize{Martial Authier}}
\author[4]{\footnotesize{Frederic Boumard}}
\author[2]{\footnotesize{Lucia Caceres}}
\author[4]{\footnotesize{Jean-François Cam}}
\author[1]{\footnotesize{Arno Claessens}}
\author[2]{\footnotesize{Samuel Damoy}}
\author[2]{\footnotesize{Pierre Delahaye}}
\author[4]{\footnotesize{Philippe Desrues}}
\author[5]{\footnotesize{Wenling Dong}}
\author[3]{\footnotesize{Antoine Drouart}}
\author[5]{\footnotesize{Patricia Duchesne}}
\author[1]{\footnotesize{Rafael Ferrer}}
\author[4]{\footnotesize{Xavier Fléchard}}
\author[5]{\footnotesize{Serge Franchoo}}
\author[2]{\footnotesize{Patrice Gangnant}}
\author[2]{\footnotesize{Sarina Geldhof}}
\author[1]{\footnotesize{Ruben P. de Groote}}
\author[2]{\footnotesize{Nathalie Lecesne}}
\author[2]{\footnotesize{Renan Leroy}}
\author[4]{\footnotesize{Julien Lory}}
\author[2]{\footnotesize{Franck Lutton}}
\author[5]{\footnotesize{Vladimir Manea}}
\author[4]{\footnotesize{Yvan Merrer}}
\author[6]{\footnotesize{Iain Moore}}
\author[2,6]{\footnotesize{Alejandro Ortiz-Cortes}}
\author[2]{\footnotesize{Benoit Osmond}}
\author[2]{\footnotesize{Julien Piot}}
\author[5]{\footnotesize{Olivier Pochon}}
\author[8,9]{\footnotesize{Sebastian Raeder}}
\author[1]{\footnotesize{Antoine de Roubin}}
\author[2]{\footnotesize{Hervé Savajols}}
\author[1]{\footnotesize{Simon Sels}}
\author[7]{\footnotesize{Dominik Studer}}
\author[10]{\footnotesize{Emil Traykov}}
\author[6]{\footnotesize{Juha Uusitalo}}
\author[4]{\footnotesize{Christophe Vandamme}}
\author[3]{\footnotesize{Marine Vandebrouck}}
\author[1]{\footnotesize{Paul Van den Bergh}}
\author[1]{\footnotesize{Piet Van Duppen}}
\author[7]{\footnotesize{Klaus Wendt}}

\affil[1]{KU Leuven, Instituut voor Kern- en Stralingsfysica, B-3001 Leuven, Belgium}
\affil[2]{GANIL, CEA/DRF-CNRS/IN2P3, B.P. 55027, 14076 Caen, France}
\affil[3]{IRFU, CEA, Université Paris-Saclay, F-91191 Gif sur Yvette, Fran}
\affil[4]{Normandie Université, ENSICAEN, UNICAEN, CNRS/IN2P3, LPC Caen, F-14000 Caen, France}
\affil[5]{IJCLab, Université Paris-Saclay, CNRS/IN2P3, IJCLab, 91405 Orsay, France}
\affil[6]{Department of Physics, University of Jyväskylä, PO Box 35 (YFL), Jyväskylä FI-40014, Finland}
\affil[7]{Institut für Physik, Johannes Gutenberg-Universität Mainz, 55128 Mainz, Germany}
\affil[8]{GSI Helmholtzzentrum für Schwerionenforschung GmbH Planckstraße 1, Darmstadt, 64291, Germany}
\affil[9]{Helmholtz Institute Mainz, Staudingerweg 18, 55128 Mainz, Germany}
\affil[10]{IPHC, Université de Strasbourg, CNRS, F-67037 Strasbourg, France}

\date{}
\begin{document}

\maketitle

\begin{abstract}
{\footnotesize    
In this paper we present the first high-resolution laser spectroscopy results obtained at the GISELE laser laboratory of the GANIL-SPIRAL2 facility, in preparation for the first experiments with the S$^3$-Low Energy Branch. Studies of neutron-deficient radioactive isotopes of erbium and tin represent the first physics cases to be studied at S$^3$. The measured isotope-shift and hyperfine structure data are presented for stable isotopes of these elements. The erbium isotopes were studied using the $4f^{12}6s^2$ $^3H_6 \rightarrow 4f^{12}(^3 H)6s6p$ $J = 5$ atomic transition (415 nm) and the tin isotopes were studied by the $5s^25p^2 (^3P_0) \rightarrow 5s^25p6s (^3P_1)$ atomic transition (286.4 nm), and are used as a benchmark of the laser setup. Additionally, the tin isotopes were studied by the $5s^25p6s (^3P_1) \rightarrow 5s^25p6p (^3P_2)$ atomic transition (811.6 nm), for which new isotope-shift data was obtained and the corresponding field-shift $F_{812}$ and mass-shift $M_{812}$ factors are presented.}
\end{abstract}

\captionsetup{justification=centering}

\begin{multicols}{2}
\section{Introduction}

Laser spectroscopy has been a powerful tool for atomic and nuclear physics research since the 1970s \cite{ambartzumian1972}. It allows measurements of atomic level energies and transition strengths, as well as nuclear ground-state properties, such as differences in mean-square charge-radii $\delta$$\langle$\textit{r}$^2$$\rangle$, magnetic dipole $\mu$ and electric quadrupole $Q$ moments, and nuclear spins \textit{I} via the effects of the electron-nucleus hyperfine interaction. Experimental atomic- and nuclear-structure information is of importance for the development of theoretical models and applications \cite{campbell2016}. However, the atomic auto-ionizing states and hyperfine structure (HFS) constants of many elements are still insufficiently known even for the stable isotopes, which makes it often difficult to find ionization schemes that are efficient and sensitive to nuclear properties. Nonetheless, laser ionization spectroscopy provides an appropriate tool for studying radioactive isotopes produced in low quantities, of the order of 1000 per second or less. An overview of the current resonance laser ionization techniques is given in \cite{marsh2014}. 

The SPIRAL2 facility in GANIL \cite{Dechery2016} is being constructed with the purpose of expanding the production of radioactive isotopes to unexplored or poorly explored areas of the nuclear chart, such as the actinide and super-heavy elements, refractory elements and isotopes along the proton drip line in the $N$ = $Z$ region around $^{100}$Sn \cite{savajols2017}, where information is scarce due to the low production rates at existing laboratories. The radioisotope production at SPIRAL2-GANIL will be driven by a superconducting LINAC accelerator. It has been designed to accelerate stable ion beams from He to U with energies from 0.75 up to 14.5 MeV/u, and intensities above 1 p$\mu$A for projectile masses up to Ni \cite{Dechery2015}. A full introduction to the facility and its scientific program is available in a recently published white book \cite{Ackermann2021}.

The accelerated heavy ions will impinge on a rotating thin target, inducing fusion-evaporation reactions. The resulting products of interest will be captured, guided and filtered by the Super Separator Spectrometer (S$^{3}$) \cite{Dechery2016} which is currently under construction. To measure ground- and isomeric-state properties of the isotopes of interest via laser spectroscopy, decay spectroscopy and mass spectrometry techniques, the S$^3$-low energy branch (S$^{3}$-LEB) was developed \cite{Ferrer2013}.   

The S$^{3}$-LEB is a gas-cell-based setup which employs the in-gas laser ionization and spectroscopy (IGLIS) technique \cite{R.Ferreretal.2017}, with the aim to selectively ionize and measure the isotope-shifts (IS) and hyperfine constants of exotic radioactive isotopes of interest in order to reveal their nuclear properties. In the IGLIS method, the reaction products are thermalized and neutralized in the buffer gas (typically purified argon at room temperature) that is constantly flowing through the gas cell at a few 100 mbar pressure. If laser ionization is performed in the gas cell, the measurement suffers from Doppler- and collisional-broadening effects of the spectral lines, to a range up to several GHz. This broadening is sufficient to mask the hyperfine splitting typically in all but the heaviest elements. To overcome this obstacle a de Laval nozzle has been installed at the exit of the gas cell to create a collimated and homogeneous hypersonic gas jet of low temperature and low density \cite{Kudryavtsev2013}, carrying the reaction products. Moving from in-gas-cell to in-gas-jet ionization allows one to perform laser spectroscopy at about an order of magnitude higher resolution without losing the high selectivity and efficiency~\cite{R.Ferreretal.2017,RFerreretal2021}. The S$^{3}$-LEB setup has been constructed and is currently being commissioned at the Laboratoire de Physique Corpusculaire Caen (LPC Caen) institute, University of Caen, France.

The SPIRAL2-GANIL primary beam characteristics together with the S$^{3}$ and S$^{3}$-LEB apparatus will facilitate the means to resolve the ground- and isomeric-state properties of exotic nuclei \cite{Ferrer2013} by means of in-gas-jet laser spectroscopy. However, one can only fully benefit from the aforementioned prospects if the laser system is capable of producing the necessary probing light for resonant excitation and ionization for the specific experimental needs \cite{Raeder2020}. Moreover, the long term wavelength and power stability of lasers, remote control of the laser wavelength and recording of these parameters is crucial for online experiments. Finally, for high resolution measurements, a single-mode (SM) pulsed laser system must be employed. 

The GISELE laser laboratory \cite{Lecesne2010} has been developed to fulfil the mentioned requirements for the S$^{3}$-LEB and consists of three types of titanium:sapphire (Ti:sa) lasers. For laser spectroscopy studies with GHz resolution suitable for the search of atomic transitions, a grating Ti:sa laser with $>$ 100 nm continuous scanning range is available. For shorter-range scans and subsequent excitation/ionization steps, a set of multi-mode (MM) birefringent filter (BRF) Ti:sa lasers can be used. Lastly, for in-gas-jet high resolution measurements an injection-locked SM Ti:sa laser system is available.

All laser systems are used to perform resonance ionization spectroscopy (RIS) measurements, in which the founding principle is to step-wise excite and ionize the valence electron of the atom. A detailed overview of the characteristics of the laser systems used for this work can be found in \cite{Mattolat2009,Moore2005,Rothe2013,Sonnenschein2017} and a more detailed description of the S$^3$-LEB setup and the first offline commissioning results are presented in \cite{romans2022}.

Here, we present narrowband laser-ionization spectroscopy measurements of stable erbium and tin isotopes. Neutron-deficient radioactive isotopes of these two elements are among the first candidates that will be studied with the S$^3$-LEB. A study of $^{152}$Er (half-life $t_{1/2}$ = 10.3 s) will be used for the online commissioning of the installation. The hyperfine structure of $^{151,151m}$Er ($t_{1/2}$ = 23.5 s, 580 ms) could also be measured in the commissioning campaign, giving access to the nuclear moments of its ground and high-spin isomeric state and bringing the nuclear structure information one-step closer to the $N$ = 82 shell closure. On the other hand, the region around the heaviest $N$ = $Z$ self-conjugate nucleus $^{100}$Sn ($t_{1/2}$ = 1.16 s), as the flagship experiment of the facility, presents a unique landscape for studying the proton-neutron interaction, nuclear shells, shapes and the mirror symmetry \cite{faestermann2013}. 

In the following, the experimental setup and measurement procedure used in this work are described. The data analysis and the results are presented in section 3 preceding the conclusion section.

\section{Methodology}
\label{ch:methodology}

The GISELE Ti:sa laser system is pumped by a single Nd:YAG pump laser (Photonics Industries DM75) operating at 532 nm and working at 10 kHz repetition rate with $\sim$160 ns pulse width. Its maximum average beam output power is 70 W. The distribution of the pump power between the different lasers is performed using polarizing beam-splitter cubes and half-wave plates. The high-resolution state scanning was performed with an injection-locked SM Ti:sa laser. This is a master-slave system, where a low power SM continuous-wave external-cavity diode-laser (ECDL) is seeded/injected into a ring amplifier Ti:sa cavity. The Ti:sa cavity has a bow-tie geometry with one of the mirrors being attached to a piezo actuator, which is controlled by a lock-in amplifier (TEM Messtechnik LaseLock) that ensures the optimum cavity length \cite{Sonnenschein2017}. As the master ECDL cavity wavelength is remotely scanned, the slave Ti:sa cavity length is adjusted by the LaseLock system to match the SM operation condition. 

Both ionization schemes used for the laser spectroscopy studies in this work are shown in Figure \ref{fig:Er Sn I RIS scheme}. The necessary seeding light for the SM narrow-band probing laser was provided by two Eagleyard gain chips (EYP-RWE-0840 and EYP-RWE-0860) installed in the ECDL (Toptica DL100). 

The emission spectrum linewidth, or laser bandwidth, $\Delta \nu$ in a seeded cavity without any dispersive elements is Fourier-limited with some additional practical limitations (fast gain medium refractive index changes and fast cavity detuning during the pump injection). The injection-locked SM Ti:sa laser provided fundamental light with a linewidth $\Delta \nu$ $\sim$ 50 MHz at full-width at half-maximum (FWHM). Such narrow $\Delta \nu$ not only allow the resolution allowing unique studies of hyperfine structure in spectra in-gas-jets, but also lead to a significant reduction in necessary output powers as the laser energy spread is so small (if the external broadening mechanisms are at the same or lower level as $\Delta \nu$). The remaining frequency steps in the ionization scheme were provided by BRF Ti:sa lasers, with a typical $\Delta \nu$ of $\sim$ 1.5-5 GHz depending on the resonator design. The second and third harmonic generation of the Ti:sa light necessary for these measurements was produced by BBO nonlinear crystals. In the case of the available SM Ti:sa laser system, the higher harmonic generation (HHG) was in single-pass geometry performed outside the cavity. In contrast, intra-cavity second-harmonic generation was performed in the BRF Ti:sa cavities. 

Using this laser system erbium atoms were resonantly excited via the $4f^{12}6s^2$ $^3H_6 \rightarrow 4f^{12}(^3 H)6s6p$ $J = 5$ transition at 415.2 nm (24 083.2 cm$^{-1}$, initially reported in \cite{Meggers1975}). The transition strength between the excited state (ES) and the ground state (GS) $A_{21}$$\,$=$\,$9.6$\times$10$^7$ s$^{-1}$ \cite{NISTweb}. The odd-even $^{167}$Er nuclear ground state (GS) spin $I$ = 7/2 \cite{NISTweb} (the $odd-even$ terminology indicates the neutron-proton numbers in the nucleus being odd or even, respectively). The second step excites the atom to an autoionizing (AI) state reported in \cite{Studer2015}, providing efficient ionization.

In the case of tin, the resonant excitation was performed via the $5p^2$ $^3P_0 \rightarrow 5p6s$ $^3P^o_1$ and $5p6s$ $^3P^o_1 \rightarrow 5p6p$ $^3P_2$ transitions at 286.4 nm and 811.6 nm, respectively (34 914.3 and 12 320.9 cm$^{-1}$). Both ES were initially reported in \cite{Meggers1975,sansonetti2005}. The transition strength from the atomic GS to the first excited state (FES) is $A_{21}$ = 5.4$\times$10$^7$ s$^{-1}$ \cite{NISTweb}, while the transition strength to the second excited state (SES) is not known. The AI state used for this work was reported in \cite{RILISweb}. All odd-even tin isotopes measured within this work have nuclear spin $I$ = 1/2 \cite{NISTweb}.

The RIS measurements are performed under vacuum inside an atomic beam unit (ABU) as described in \cite{VINCK2014}. The setup consists of a cross chamber, housing an oven, a set of apertures and an electrode assembly, a time-of-flight (TOF) section and a micro-channel plate (MCP) detector. To produce the erbium and tin atoms, a solution of Er$_2$O$_3$ in 5 \% HNO$_3$ and Sn in 10 \% HCl was dried on a tantalum foil and placed inside the ABU oven that was resistively heated to a temperature $T >$ 1000$\degree$ C. An aperture and electrode assembly is used to reduce the divergence of the vertically diffusing atomic plume to 2.6$^{\circ}$ opening angle at the laser-atom interaction zone. The multiple laser beams are overlapped with polarizing beam optics and dichroic mirrors. For a temperature of 1500$^{\circ}$ C and the used setup geometry, the maximum total (Doppler + natural) width of the atomic plume is $\sim$ 73 MHz in the case of the excitation transition of erbium atoms. With a slightly lower necessary evaporation temperatures of 1250$^{\circ}$ C for tin the total (Doppler + natural) width  is $\sim$ 116 MHz for the FES transition (for the SES transition an estimate could not be made as A$_{21}$ is not known, but a lower limit due to Doppler width is 41 MHz). To further minimize Doppler width and maximize the photon flux density, the laser beams are focused in the photon-atom interaction area by telescopes to $\sim$ 1-2 mm diameter size.  

The photon-atom interaction region is located in between two acceleration electrodes typically biased to electrostatic potentials of $\sim$ 1.3 kV and 1 kV, plus an additional electrode at ground potential. The three electrodes accelerate the photo-ions towards the MCP detector through a time-of-flight (TOF) section of $\sim$ 50 cm length. The MCP detector is biased at -2 kV to attract the photo-ions and convert their arrival into an electric signal. This analog signal is amplified and converted by a constant-fraction discriminator to a logic (NIM) signal. The digital signal is then fed into a time-to-digital converter (TDC) triggered by the pump laser pulses with a maximum temporal resolution of 4 ns. This allows both to count ions and to measure their TOF.

The timing between the laser pulses of the corresponding ionization schemes was monitored in an oscilloscope using photo-diodes detecting a 10 $\%$ fraction of the light reflected either from the cavity BRF, or a beam sampler placed in the laser output beam. The remaining 90 $\%$ fraction of BRF/beam sampler reflected light was sent to the HighFinesse WS 7 wavemeter (absolute accuracy 30 MHz or 20 $\%$ of the $\Delta \nu$, whichever is larger). The calibration of the wavemeter was performed with a frequency-stabilized He:Ne laser (Thorlabs HRS015B). The longitudinal mode-structure of the SM Ti:sa output light was observed with a Toptica FPI 100-0750-3V0 scanning Fabry-Perot interferometer (SFPI) with 1 GHz free spectral range and typical finesse of 500.

\begin{Figure}
    \centering
    \includegraphics[trim=0.05cm 0.25cm 0.2cm 0.1cm,clip,width=\linewidth]{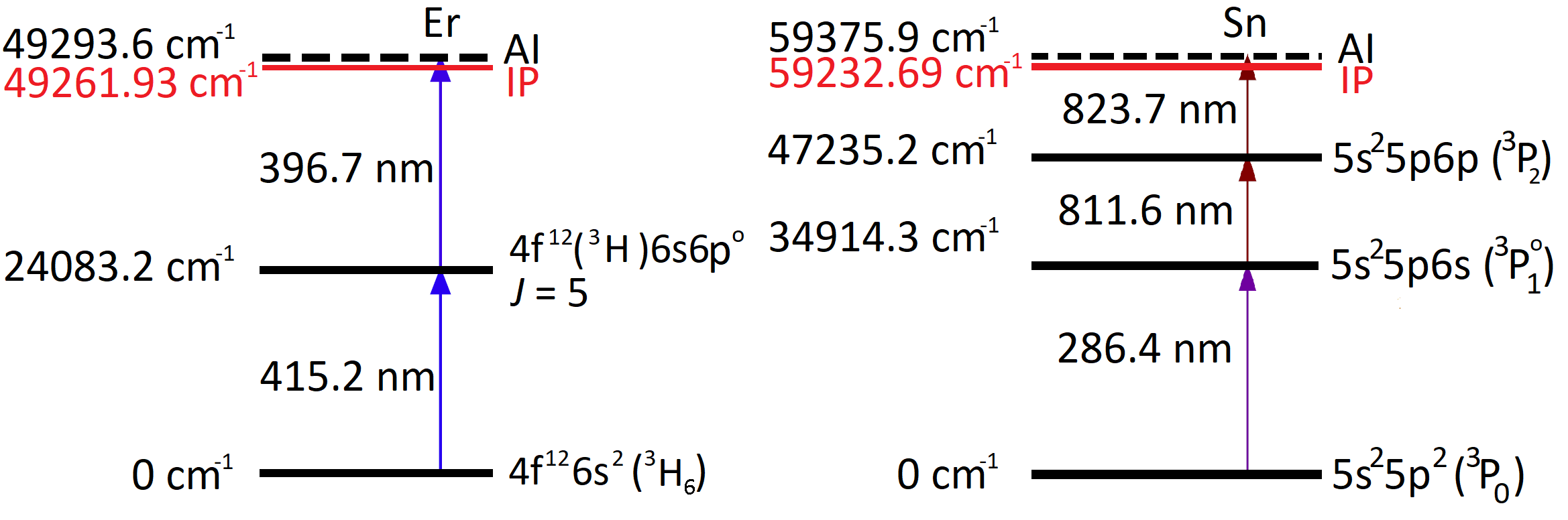}
    \captionof{figure}{(Left) Er two-step and (right) Sn three-step ionization schemes used for RIS measurements in this work \cite{Studer2015,RILISweb}. On the left hand side of each ionization scheme diagram, the energies of the excited states, the ionization potential (IP) and the populated auto-ionizing (AI) state are given in cm$^{-1}$. On the right hand side the electron configuration is shown. The electron configuration and energy levels are taken from \cite{NISTweb}}
    \label{fig:Er Sn I RIS scheme}
\end{Figure}

\section{Results}
\label{ch:results}

\subsection{Data analysis}

Prior to the laser-spectroscopy measurements, power broadening tests were carried out for the excitation step used for scanning the wavelength in order to establish a compromise between acceptable count rates and minimized power broadening of the resonance signal. The saturation power for the transition to the the FES in erbium was established to be 145(40) $\mu$W \cite{romans2022} and for the FES and the SES in tin were measured to be 427(40) $\mu$W and 7.9(5) mW, respectively. These values are obtained for the quoted beam diameter sizes of 1-2 mm. The saturation power tests in the case of erbium were performed with SM Ti:sa laser, while for tin the BRF Ti:sa lasers were used.

In the case of erbium, to reduce the contribution of the power broadening to the spectral linewidth as much as possible, spectroscopy scans were taken with the FES powers $\leq$ 20 $\mu$W. This resulted in an average FWHM of 140 MHz, which could be explained by a combination of the SM Ti:sa laser second harmonic linewidth of $\leq$ 100 MHz and additional broadening due to temporally synchronized excitation and ionization RIS steps. This is roughly twice the expected Doppler width of the excited atom ensemble, but due to the reduction in count rates, the 10 - 20 $\mu$W range was used for the FES in the measurements. 

In the case of tin, while performing spectroscopy on the FES transition at a saturation power $P_0$, the observed FWHM values of the individual isotope resonances were $\sim$ 150 MHz, which is expected from the SM Ti:sa system after third harmonic generation and it is close to the approximated Doppler width. Therefore, no additional check of the lower powers was performed as the resonance FWHM corresponded to our light source linewidth. However, to obtain optimum count rates our measurements were carried out with the second step power of $\sim$ 135 mW. This large increase in second step power was needed once the SM Ti:sa laser was used for measurements as the initial tests were carried out with the BRF Ti:sa lasers. The use of the SM Ti:Sa laser resulted in a temporal displacement of the three pulses involved in the ionization scheme that could not be optimally synchronized and the corresponding signal loss could be compensated by increasing the laser power of the second step.

With this setup the laser frequency of the scanning laser was varied and the counts in the TDC spectrum were recorded. A resulting summed TOF spectrum for all the FES scan steps with the ionization step set on resonance is presented in Figure \ref{fig:Er & Sn I RIS tof}. The resulting ABU TOF mass resolving power for erbium and tin are determined to be $R = \mathrm{TOF}/(2 \times \mathrm{FWHM}) \sim 260$ and $\sim 225$, with TOF = 21.4 and 13.5 $\mu$s, respectively. In Figure \ref{fig:Er & Sn I RIS tof} the TOF origin is offset with respect to the laser pulse by an internal TDC delay. 

By plotting the obtained signal as function of the excitation laser frequency within a specified TOF gate, the optical spectra of each isotope are obtained, for which the analysis will be discussed in the following. The typical width of the ion signals in the MCP were found to be 35 ns. Typical TOF gates ranged from 20 to 80 ns width, with larger gates chosen for the isotopes that are fully separated from the neighboring ones. 

The individual isotope hyperfine spectra are fitted with the SATLAS package \cite{GINS2018} using the \textit{chisquare fit} method and Voigt profile description. For an appropriate uncertainty estimation of the fits, the amount of resonance peak with the background points of erbium and tin data sets has been reduced to span over a $\sim$ 400 - 500 MHz frequency range. The exception were the HFS spectra in tin, where the 1000 MHz range was used.

The uncertainty of the number of counts for a given scan point is obtained assuming a Poisson distribution of the events $\sigma_N$ = $\sqrt{N}$. Following the SATLAS fit of each species in the spectrum, the statistical uncertainty of each centroid is multiplied by the square-root of reduced $\chi^2$, if reduced $\chi^2$ $\geq$ 1. For each scan, the extracted IS to the reference isotope is calculated by subtraction of the fitted centroids. The final IS values are computed as weighted averages (IS$_{\mbox{\tiny{WA}}}$) of the individual ones. Because the reduced $\chi^2$ computed with the individual IS values is greater than 1, the standard deviation is calculated from square-root of the sum of individual squared centroid uncertainties, which is used as the final uncertainty of our weighted average results. Notice that in such laser spectroscopy measurements low average laser intensities ($\sim$ 1 W/cm$^2$) with peak intensities of 2-10 kW/cm$^2$, from pulse widths of 10-50 ns, in the IR - UV emission spectrum range are used, which do not lead to an observable AC Stark shift with respect to the centroid uncertainty range.

\begin{Figure}
    \centering
    \includegraphics[trim=0.35cm 0.1cm 0.2cm 0cm,clip,width=\linewidth]{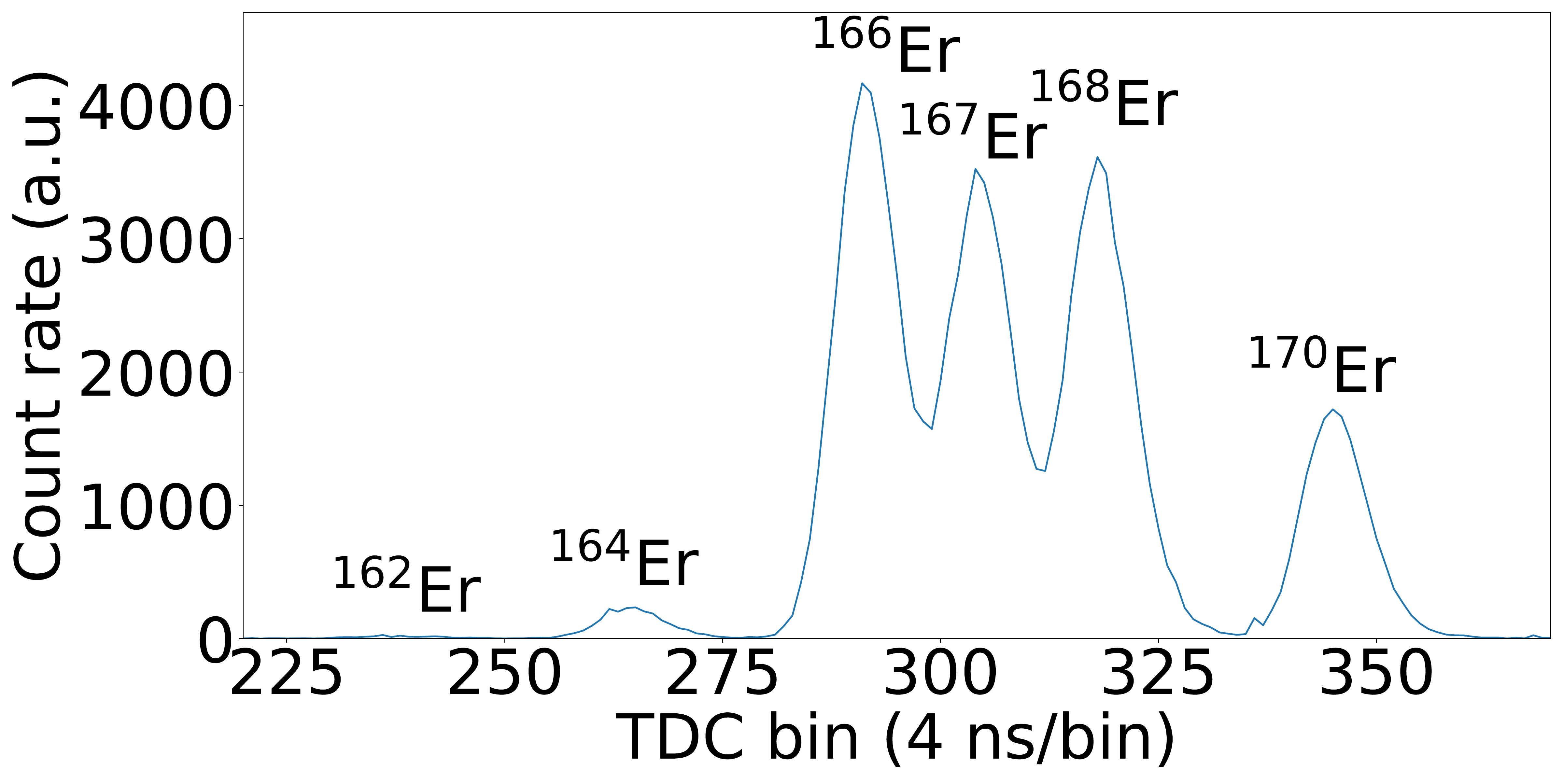}
    \includegraphics[trim=0.35cm 0.5cm 0.2cm 0cm,clip,width=\linewidth]{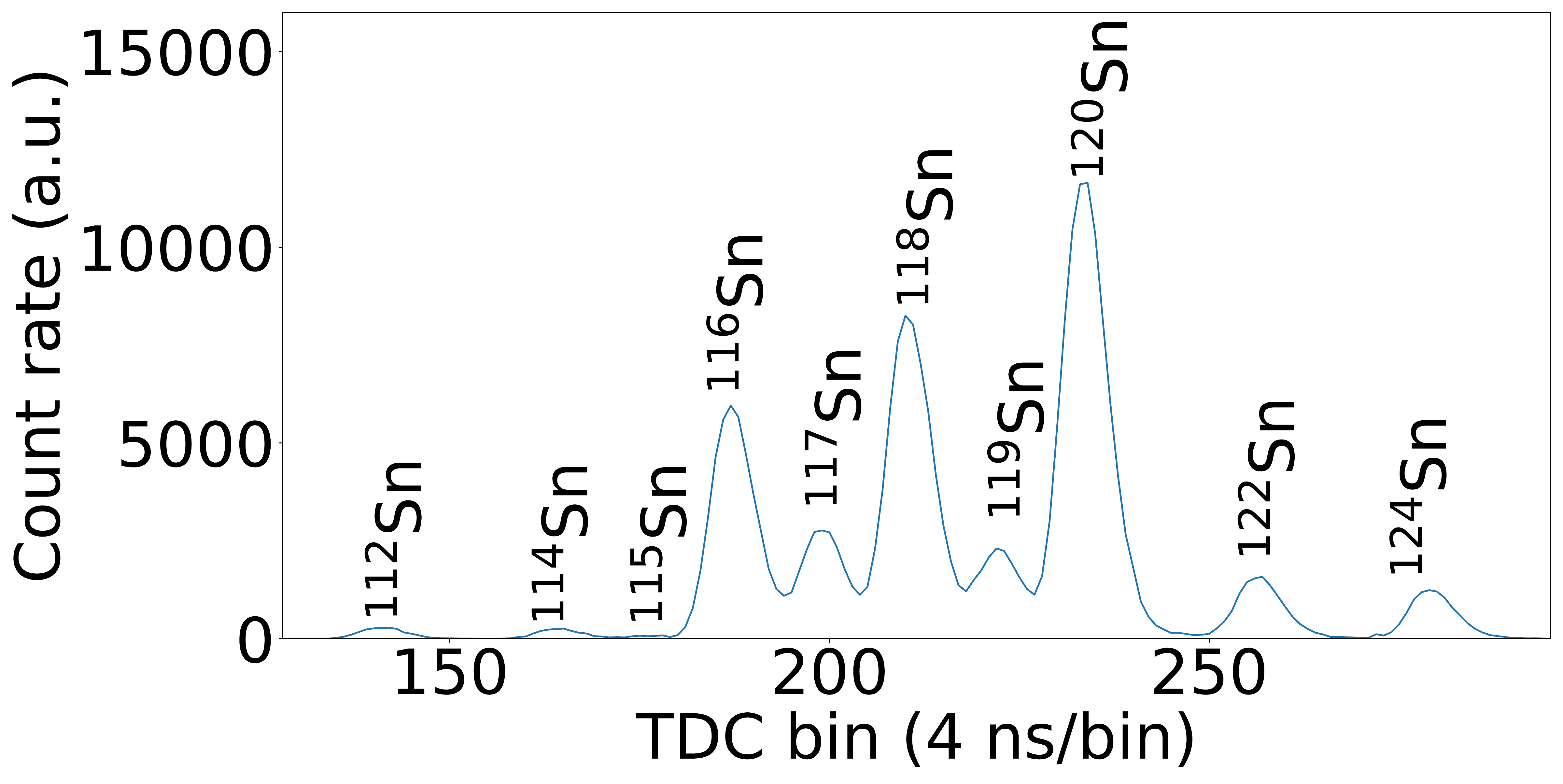}
    \captionof{figure}{Cumulated TOF spectrum of stable erbium (top) and tin (bottom) ions following a RIS scan using the schemes shown in Figure \ref{fig:Er Sn I RIS scheme}. The peak intensities reflect the natural abundance of the stable isotopes.}
    \label{fig:Er & Sn I RIS tof}
\end{Figure}

The IS can be related to the change in nuclear mean-square charge radius $\delta \langle r^2\rangle^{A^{\prime},A}$ = $r^{A^{\prime}}$ - $r^{A}$ between the measured, $A^{\prime}$, and a reference, $A$, isotope via approximation (represented for the 415 nm transition in Er)

\begin{equation}
    \delta\nu_{415}^{A^{\prime},A} = F_{415}\delta\langle r^2\rangle^{A^{\prime},A} + M_{415}\frac{A^{\prime}-A}{A^{\prime}A},
    \label{eq:IS}
\end{equation}
where $F$ and $M$ are the atomic field- and mass-shift factors for the specific transition. The $M$ factor can be further split into the normal mass-shift constant contribution $K_{\mbox{\tiny{NMS}}}$, describing the change in the motion of the center of mass within the framework of the uncorrelated individual electrons, and the specific mass-shift constant $K_{\mbox{\tiny{SMS}}}$, describing the change in a multi-electron system correlated motion. Typically, the specific mass-shift factors, and thus $M$ factors, are not known and need to be calculated based on atomic theory. An alternative way to extract these variables is via King plot analysis with data from reference IS measurements obtained with a different transition \cite{king2013}.

The atomic factors from our results can be obtained by multiplying Eq. \ref{eq:IS} for both the reference transition and the one used in the present work by the modification factor $\mu^{A^{\prime},A}$ = $A^{\prime}A/(A^{\prime}-A)$ (calculated from \cite{Wang2021}) and after eliminating the $\delta \langle r^2\rangle^{A^{\prime},A}$ term between the two equations, one obtains the following expression which relates the $F$ and $M$ factors of the two transitions:

\begin{equation}
    \mu^{A^{\prime},A}\delta\nu_{\mbox{new}}^{A^{\prime},A} = \frac{F_{\mbox{new}}}{F_{\mbox{ref}}}\mu^{A^{\prime},A}\delta\nu_{\mbox{ref}}^{A^{\prime},A} + M_{\mbox{new}} - \frac{F_{\mbox{new}}}{F_{\mbox{ref}}}M_{\mbox{ref}}.
    \label{eq:King plot}
\end{equation}

\subsection{RIS of Er I}
\label{ss:Er I ris}

\begin{figure}[h!]
    \centering
    \includegraphics[trim=1.8cm 0.7cm 0.6cm 0.45cm,clip,width=0.93\linewidth]{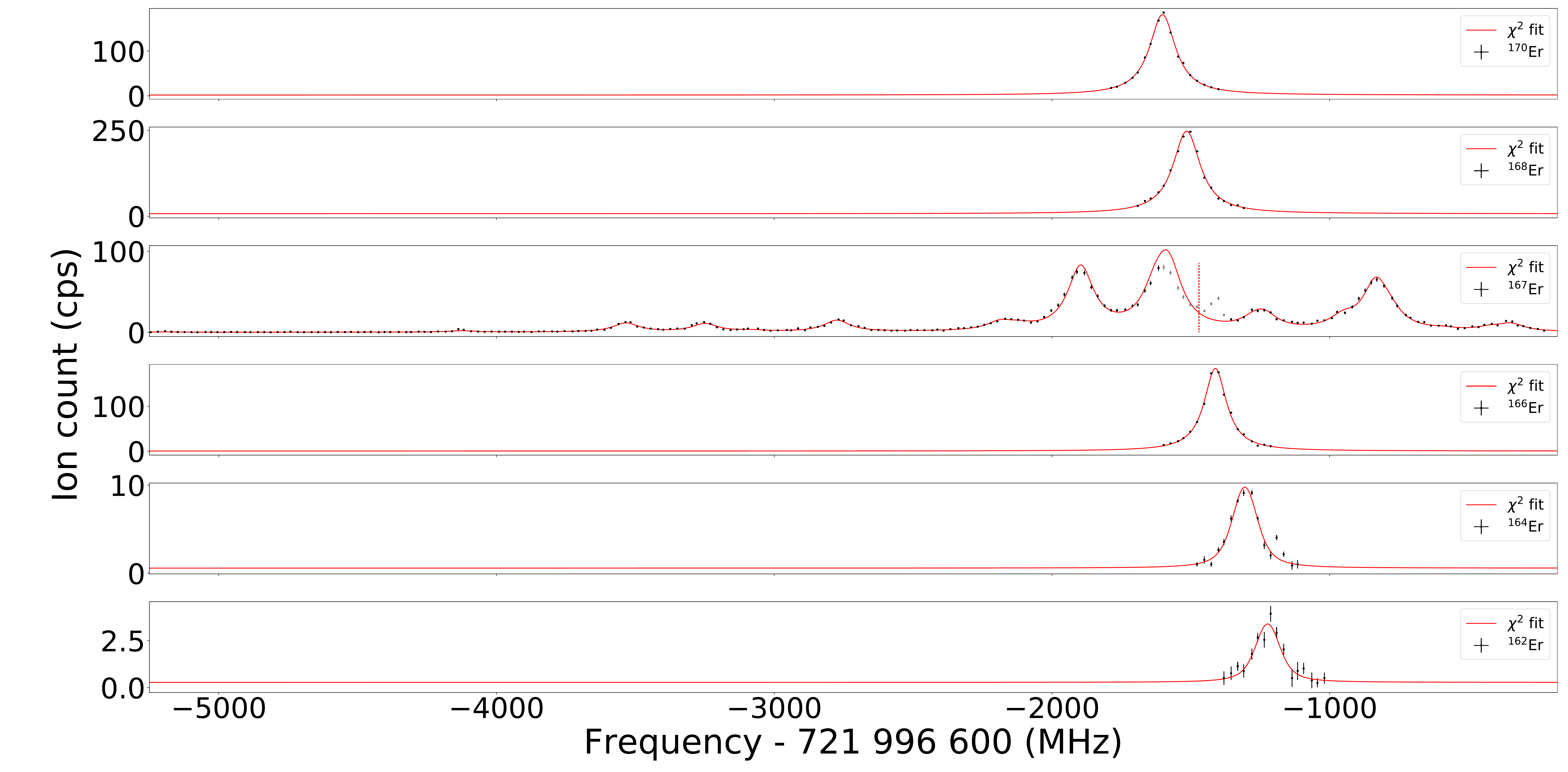}
    \includegraphics[trim=0.5cm 0.4cm 0.35cm 0.35cm,clip,width=0.93\linewidth]{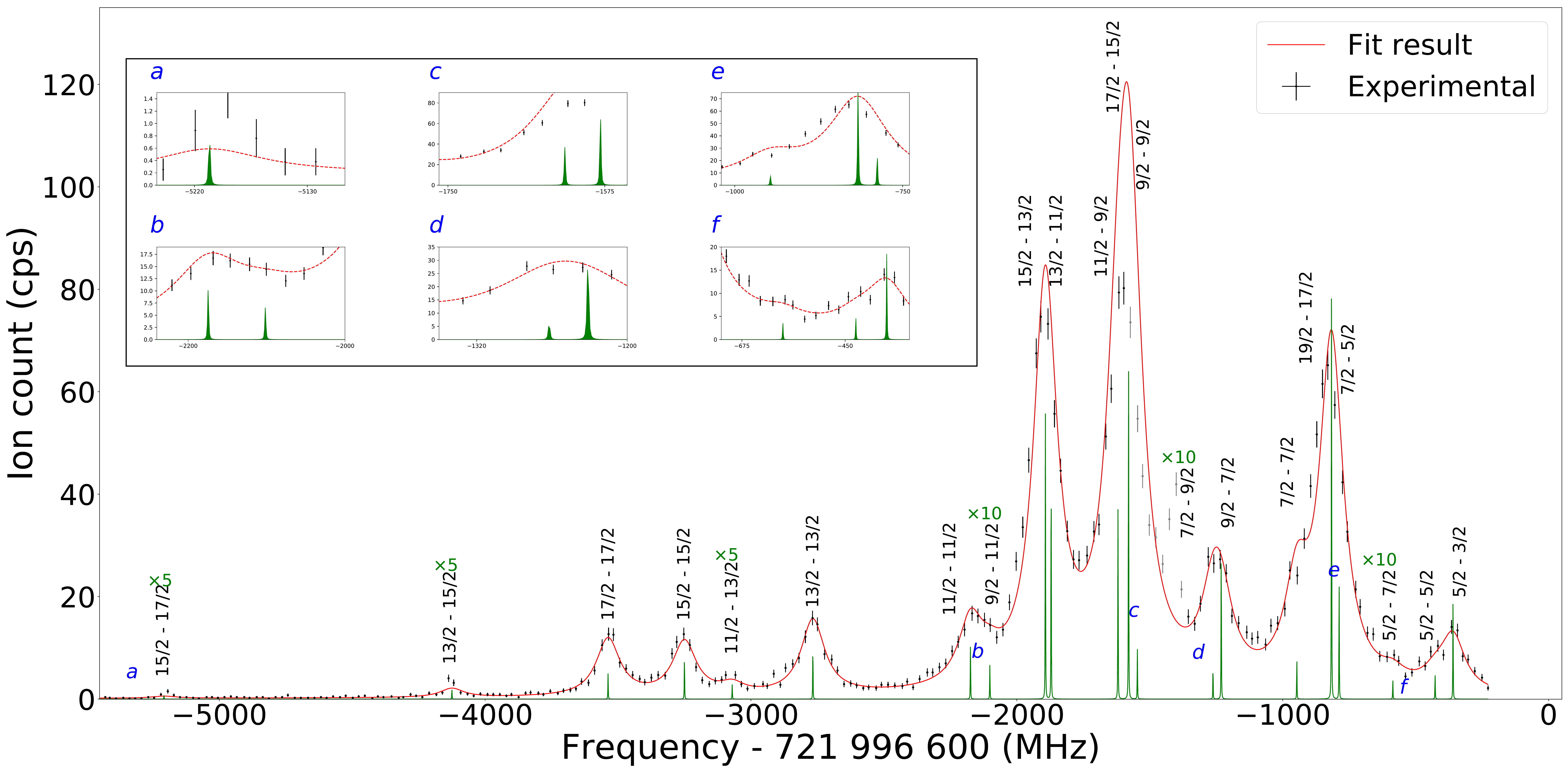}
    \caption{(Top) $^{170-162}$Er RIS scan data shown in black of the $4f^{12}6s^2$ $^3H_6$ $\rightarrow$ $4f^{12}(^3 H)6s6p$ $J$ = 5 atomic transition. Data has been fitted with the SATLAS \cite{GINS2018} \textit{chisquare fit} method (red). (Bottom) $^{167}$Er spectrum from the top panel overlapped with expected HFS peak positions (green) from atomic theory calculations using provided $I$ and $J$ values and fit parameters $A$, $B$ and centroid. Each HFS peak is labeled by the corresponding transition from the ground state $F$ to the excited state $F^{\prime}$ HFS levels. The green peak FWHM are fixed to the resulting Voigt profiles from FWHM$_{\mbox{\tiny{Gauss}}}$ and FWHM$_{\mbox{\tiny{Lorentz}}}$ of 250 kHz, to clearly indicate the individual HFS peaks. The weakest or unresolved multiplet hyperfine structure peaks have been highlighted in the top left box. For visualization purposes the least intense expected (green) individual HFS peaks are magnified by a factor of 5 or 10, indicated by $\times$5 and $\times$10 above the specific component, respectively. Grey data points for the $^{167}$Er spectrum are not included in the fit analysis due to mass contamination influence from $^{166,168}$Er. Measurements were carried out at excitation and ionization steps of 10 $\mu$W and 70 mW. See text for details.}
    \label{fig:Er I RIS}
\end{figure}

An example of the spectra measured with an optimum power of the excitation step of 10 $\mu$W and the ionization step of 70 mW for the $^{170-162}$Er isotopic chain is presented in the top panel of Figure \ref{fig:Er I RIS}. The red dotted line in the $^{167}$Er spectrum represents the fitted centroid value. 

In the bottom panel of Figure \ref{fig:Er I RIS} the $^{167}$Er spectrum is shown in more detail, where 21 HFS peaks illustrated by the green lines are expected due to the high total angular momentum $J$ values of 6 and 5 for the $4f^{12}6s^2$ $^3H_6$ and $4f^{12}(^3 H)6s6p$ $J$ = 5 states and nuclear GS spin $I$ of 7/2. In our measurement represented by the black data points all hyperfine structure (HFS) peaks have been measured and fitted with SATLAS indicated by the red curve. To extract the HFS $A$ and $B$ constants, all variables were left as free parameters. As the doublets and triplets of the HFS could not be fully resolved, for all but one HFS triplet the individual peak intensities were also left as free parameters. The exception case is described below.

The highlighted grey data points in the $^{167}$Er spectra of Figure \ref{fig:Er I RIS} in the vicinity from -1600 to -1300 MHz with respect to $\nu_0$ = 721 996.6 GHz, overlapping with the $^{166}$Er and $^{168}$Er resonances, have been excluded from the fitting procedure as they are influenced by a mass contamination. The influence can be deduced from the TOF spectra (see Figure \ref{fig:Er & Sn I RIS tof}) and the $^{166,168}$Er spectra (see the top panel of Figure \ref{fig:Er I RIS}). The contamination influence was further tested by shifting the selected $^{167}$Er TOF window closer or further from the $^{166}$Er and $^{168}$Er peak respectively. As the grey exclusion area partially overlaps with $^{167}$Er HFS triplet containing 11/2-9/2, 17/2-15/2 and 9/2-9/2 peaks, the relative intensities of the three peaks have been fixed using the corresponding Racah coefficients. This procedure is necessary to avoid that the $\chi^2$-minimization fitting method artificially increases the HFS component intensity because of the lack of data within the grey exclusion area.\\ \\ \\ \\ \\ \\

In total 20 scans were recorded and analysed following the procedure described in the previous section, in all of them the IS between the even-even isotopes could be extracted, while only in 11 scans the HFS of $^{167}$Er could be fitted. The weighted average (WA) of IS and HFS constants were calculated. To the best of our knowledge, the IS values for this transition have not been published before. Ground- and excited-state HFS coefficients $A$ and $B$ obtained in this work are in agreement with the literature as presented in Table$\,$\ref{tab:Er I RIS IS+HFS}. The presented $\sigma$($\Delta\nu$) are combined statistical and systematic uncertainties. The statistical uncertainty is obtained from the fitting procedure. The systematic uncertainty value of 6 MHz is taken from \cite{verlinde2020}, where a slightly worse absolute accuracy wavemeter from the same manufacturer working within the same wavelength range as in these measurements has been characterized.  

\begin{figure}[h!]
    \centering
    \includegraphics[trim=0.1cm 0.1cm 0.1cm 0.1cm,clip,width=0.5\linewidth]{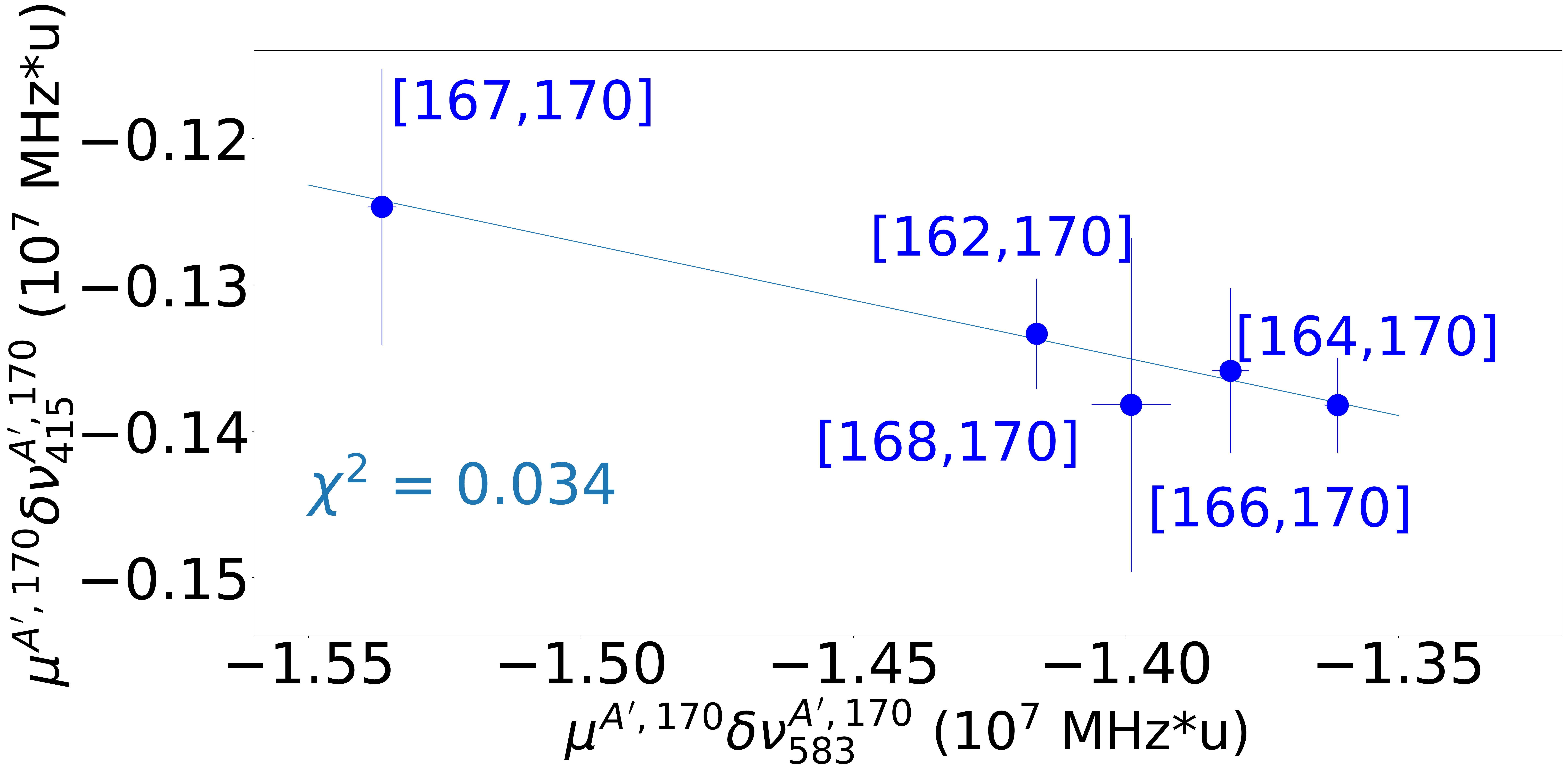}
    \includegraphics[trim=0.1cm 0.1cm 0.1cm 0.1cm,clip,width=0.47\linewidth]{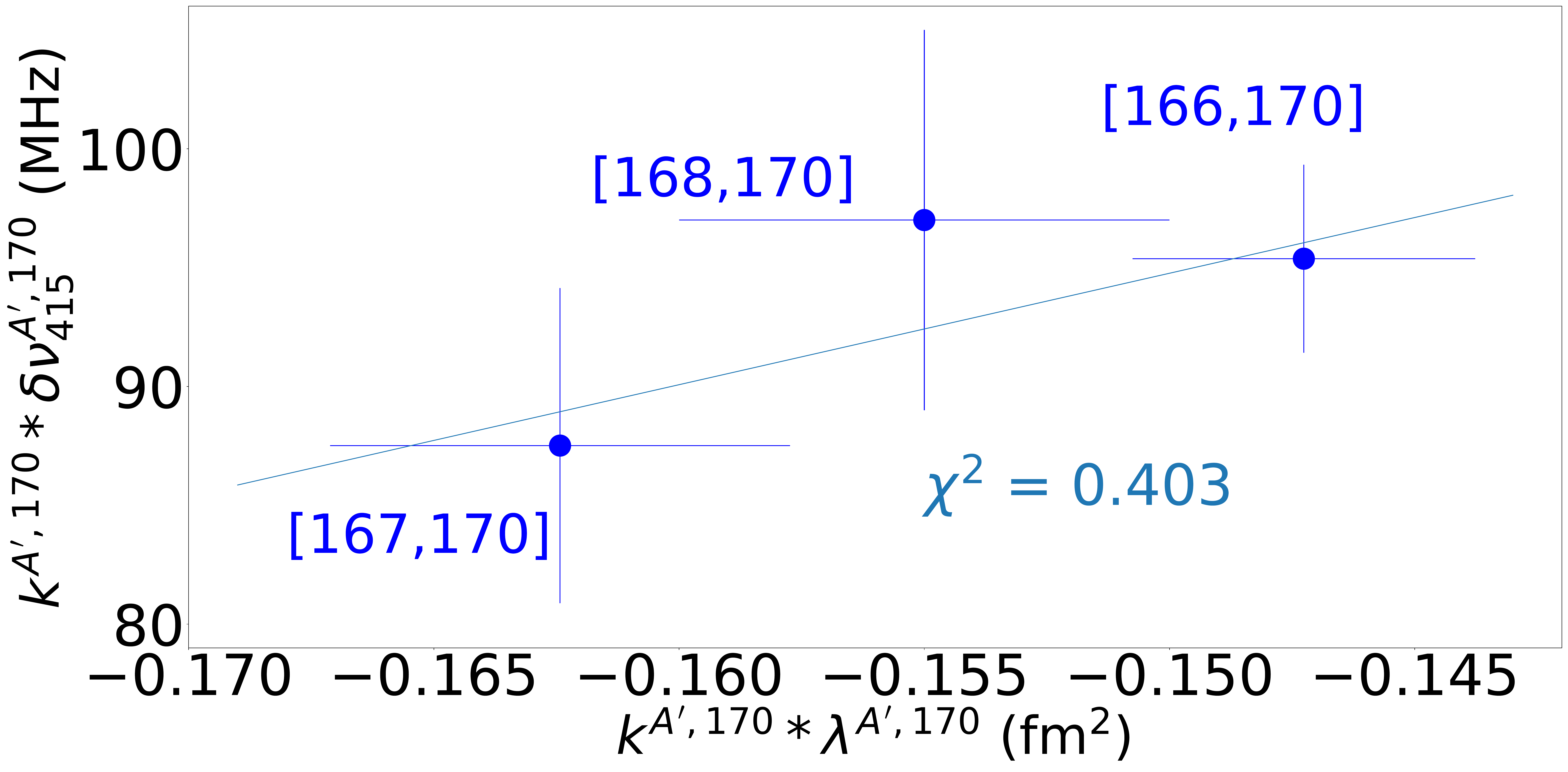}
    \caption{King plot of modified IS of 415 nm transition as a function of modified (left) IS of 582.7 nm transition from \cite{okamura1987} and and (right) muonic X-ray $\lambda^{A^{\prime},170}$ data from \cite{FrickeHeilig2004}.}
    \label{fig:King plot Er}
\end{figure}

In order to verify the reliability of our results via the linearity of the King plot analysis, we are using the $4f^{12}6s^2$ $^3H_6 \rightarrow$ $4f^{12}$($^3H_6$)$6s6p$ $^3P_1$ atomic transition at 582.7 nm as our IS measurement reference from \cite{okamura1987} and $F_{583}$ = -8.08$\,$GHz/fm$^2$ and $M_{583}$ = 282$\,$GHz*u factors from \cite{otten1989,Angeli2013}. No uncertainties were provided for $F_{583}$ and $M_{583}$. We consider the Seltzer correction accounting for higher radial moments to be included in the $F$ factor value \cite{Seltzer1916}. The obtained King plot is presented in the left panel of the Figure \ref{fig:King plot Er}. The errorbars for the individual points are amplified by the modification factor $\mu^{A^{\prime},A}$ multiplication. The reduced $\chi^2$ value for the linear fit is indicated. From the linear fit of data represented by Eq. \ref{eq:King plot}, the slope $\frac{F_{415}}{F_{583}}$ = -0.0787(95) and intercept $M_{415} - \frac{F_{415}}{F_{583}}M_{583}$ = -2452(133)$\,$GHz*u, respectively. 

However, the lack of uncertainty estimates of the $F_{583}$ and $M_{583}$ factors prevent a reliable extraction of the $F_{415}$ and $M_{415}$ factors. For this purpose the available muonic X-ray data of the parameter $\lambda^{A^{\prime},A}$ data from \cite{FrickeHeilig2004} have been used, where the nuclear parameter $\lambda$ = $\delta \langle r^2\rangle$ + $C_1 \delta \langle r^4\rangle$ + $C_2 \delta \langle r^6\rangle$ + $\dots$ \cite{angeli2004} encompasses the relative size of the nucleus. The obtained King plot in this case is presented in the right panel of the Figure \ref{fig:King plot Er}. In this case the $F_{415}$ and $M_{415}$ factors for the 415 nm transition along with their uncertainties are obtained directly from the linear fit using Eq. \ref{eq:IS} multiplied by a modification factor $k^{A^{\prime},A}$ = $A^{\prime}A/(A^{\prime}-A)$ $\times$ $(A^{\prime}-A)/A^{\prime}A$. The $F_{415}$ factor is obtained directly from the fit slope, and the intercept provides $\frac{A^{\prime}-A}{A^{\prime}A}$ $\times$ $M_{exp}$. Therefore, $M_{415}$ = slope/$\frac{A^{\prime}-A}{A^{\prime}A}$. From the linear fit both factors have been determined to be 

\begin{align*}
    F_{415} = 469(322) \: \mathrm{MHz\ fm^{-2}} \mathrm{and} \\
    M_{415} = -2352(698) \: \mathrm{GHz*u}. \\
\end{align*}

\end{multicols}

\begin{table}[h!]
    \footnotesize
    \centering
    \begin{tabular}{c c c || c c c c c c}
        \hline
        \hline
         \multicolumn{3}{c ||}{$\Delta\nu_{\mbox{WA}}^{A^{\prime},170}$ (MHz)} & \multicolumn{6}{c}{$^{167}$Er HFS coefficients} \\
         
         \multicolumn{3}{c ||}{$4f^{12}6s^2$ $^3H_6 \rightarrow 4f^{12}(^3 H)6s6p$ $J$ = 5} & \multicolumn{3}{c ||}{$4f^{12}6s^2$ $^3H_6$} & \multicolumn{3}{c}{$4f^{12}(^3 H_5)6s6p$ $J$ = 5} \\
         \hline 
         & & \multicolumn{1}{c ||}{} & \multicolumn{3}{c ||}{} & & &  \\
         
         A & $I^{\pi}$ & \multicolumn{1}{c ||}{This work} & Ref. & \multicolumn{1}{c}{$A$ (MHz)} & \multicolumn{1}{c ||}{$B$ (MHz)} & Ref. & $A$ (MHz) & $B$ (MHz) \\

        168 & $0^+$ & 97(8) & & & \multicolumn{1}{c ||}{\multirow{2}{5em}{}} & & &  \\
        
        167 & 7/2$^+$ & 132(10) & This work &  -121.80(75) & \multicolumn{1}{c ||}{-4563(53)} & This work & -147.66(83) & \multicolumn{1}{c}{-1888(58)} \\ 
        
        166 & $0^+$ & 193(8) &\multirow{1}{2em}{\cite{Childs1983}} & \multirow{1}{6em}{\, -120.487(1)} & \multicolumn{1}{c ||}{\multirow{1}{7em}{\: -4552.984(10)}} & \multirow{1}{2em}{\cite{SAAhmad1985}} & \multirow{1}{5em}{\; -146.6(3)} & \multirow{1}{5em}{\ -1874(16)} \\
        
        164 & $0^+$ & 298(7) & \multirow{1}{2em}{\cite{okamura1987}} & \multirow{1}{6em}{\quad -120.8(3)} & \multicolumn{1}{c ||}{\multirow{1}{6em}{\quad -4546(11)}} & & &   \\

        162 & $0^+$ &\ 388(11) & \multirow{1}{2em}{\cite{Frisch2013}} & \multirow{1}{4em}{\ -120.42} & \multicolumn{1}{c ||}{\multirow{1}{3em}{ -4554}} & & & \\
        \hline
        \hline
    \end{tabular}
    \caption{Extracted weighted average IS values $\Delta\nu_{\mbox{WA}}^{A^{\prime},170}$ = $\nu_{\mbox{WA}}^{A^{\prime}}$ - $\nu_{\mbox{WA}}^{170}$ of stable $^{168-162}$Er isotopes with respect to $^{170}$Er and hyperfine structure $A$ and $B$ coefficients for the $4f^{12}6s^2$ $^3H_6$ and $4f^{12}(^3 H)6s6p$ $J$ = 5 atomic states of $^{167}$Er. The $\sigma(\Delta\nu_{\mbox{WA}}^{A{\prime},170})$ indicated in parenthesis represents the combined statistical and systematic uncertainties. The $\sigma(A)$ and $\sigma(B)$ indicated in parenthesis is the standard deviation of the data set. See text for details.}
    \label{tab:Er I RIS IS+HFS}
\end{table}

\begin{multicols}{2}

\end{multicols}

\begin{multicols}{2}

\subsection{RIS of Sn I}

\begin{figure}[h!]
    \centering
    \includegraphics[trim=3.8cm 0cm 4.8cm 1.2cm,clip,width=0.96\linewidth]{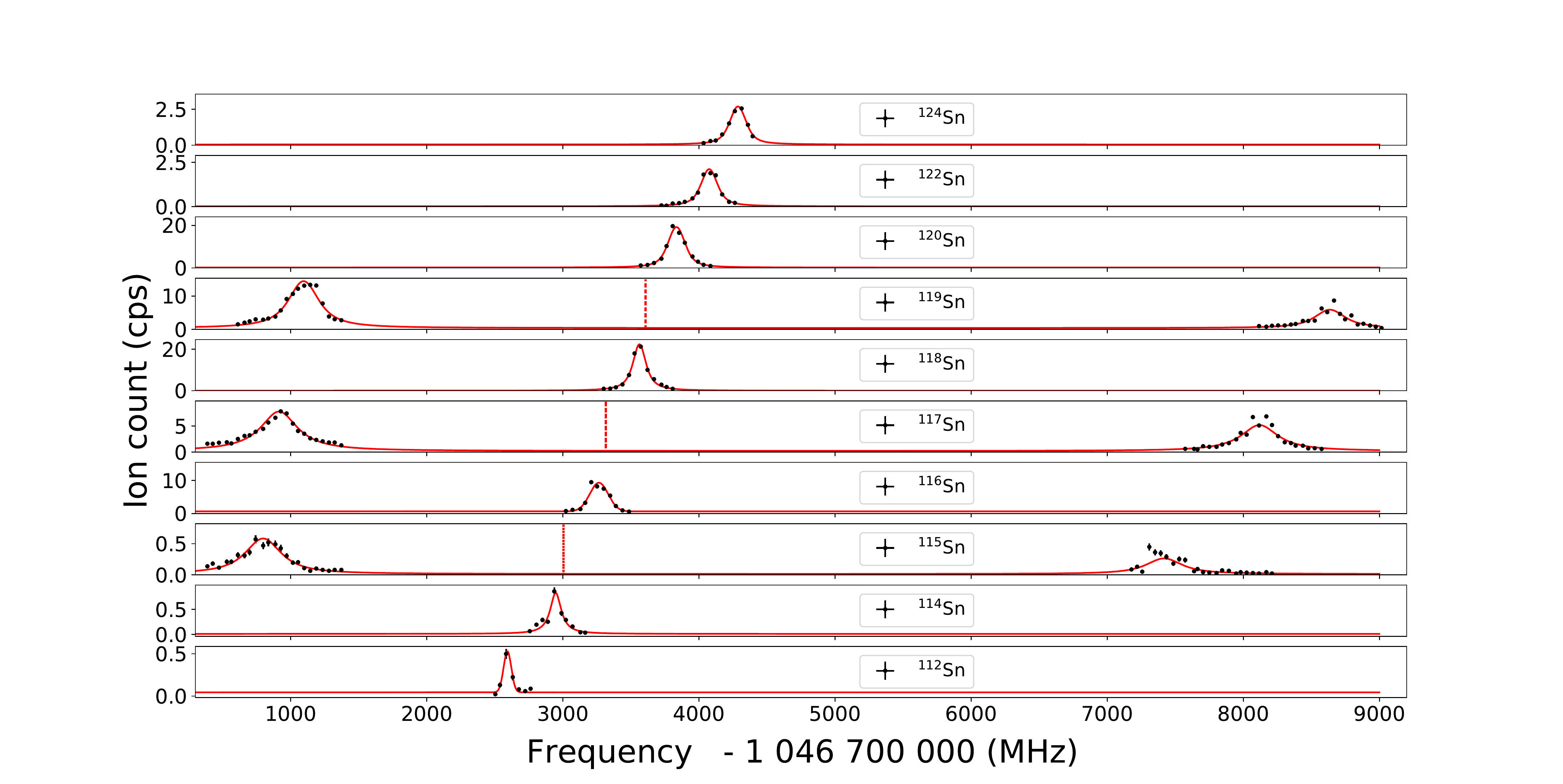}
    \includegraphics[trim=1.5cm 0.6cm 0.5cm 0.1cm,clip,width=0.96\linewidth]{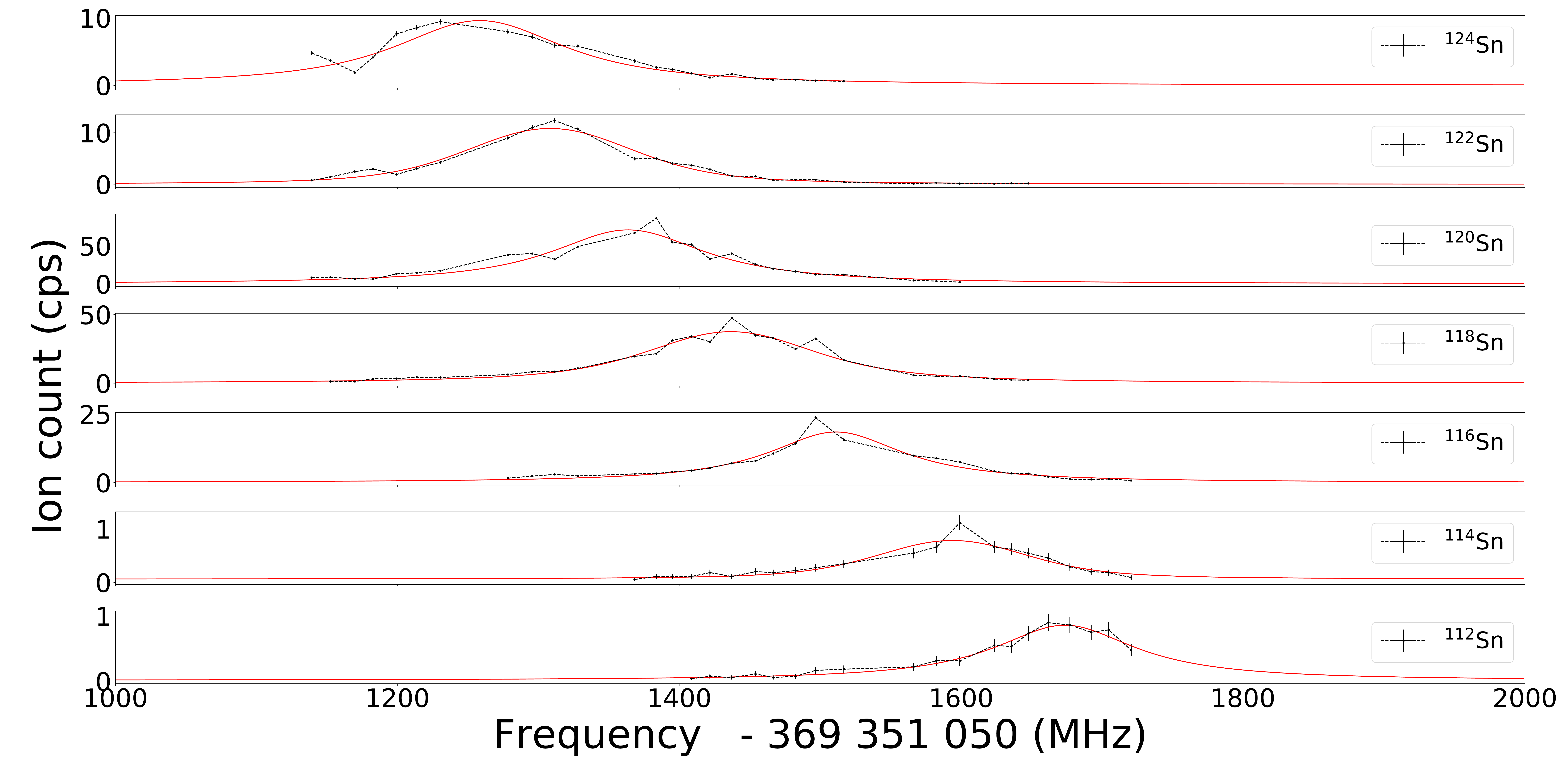}
    \caption{$^{124-112}$Sn RIS scan data shown in black. The data has been fitted with SATLAS \cite{GINS2018} \textit{chisquare fit} method (red) of the (top) FES $5p^2$ $^3P_0 \rightarrow 5p6s$ $^3P^o_1$ (286.4 nm) and (bottom) SES $5p6s$ $^3P^o_1 \rightarrow 5p6p$ $^3P$ (811.6 nm) atomic transitions. Measurements for the FES probing were carried out at $P_1$ = 180 $\mu$W, $P_2$ = 135 mW and $P_3$ $\sim$ 50 mW, and for the SES probing at $P_1$ $\sim$ 10 mW, $P_2$ = 0.5 mW and $P_3$ $\sim$ 550 mW.}
    \label{fig:Sn RIS}
\end{figure}

The Sn I RIS scans for the transitions of the FES and the SES are shown in the top and bottom panels of Figure \ref{fig:Sn RIS} with black data points. Attention must be paid to differing scanned frequency scales for the SES versus the FES probing, which reveals a significant reduction in the IS magnitude. The fitting is indicated by the red curve and in the case of odd-even isotopes the red dotted lines present the fitted centroid value. The odd isotopes were not measured with SES and for the FES one expects two HFS components due to the selection rules of total angular momentum $F$ \cite{NISTweb}. For the fitting procedure the individual peak intensities were left as free variables as isotope spectra could not be measured in one continuous scan and intensity variations of the atom source were present. 

While probing the transition of the SES a sensitivity to the MM structure of the broad-band BRF Ti:sa with (third harmonic light) linewidth of $\sim$ 15 GHz that was probing the FES transition became obvious. As the tin atoms interact with the photons from the MM laser with mode-spacing in the third harmonic of $\sim$ 900 MHz \cite{Rothe2013,herrera2013}, the scanned resonance of the SES using the SM Ti:sa laser suffered from intensity fluctuations due to these modes. However, these issues did not prohibit extracting IS from both the FES and the SES and HFS constants from the FES. The spectroscopy scans of both states were analyzed and the result uncertainties $\sigma$($\Delta\nu_{\mbox{WA}}^{A{\prime},124}$) and $\sigma$($\Delta\nu_{\mbox{WA}}^{A{\prime},122}$) were calculated following the procedure described earlier (see Section \ref{ss:Er I ris}).

For the FES probing, a total number of 23 scans were recorded. Out of these, nine allowed the extraction of the IS between the high abundance isotopes $^{116,118,120}$Sn, seven for $^{122}$Sn, five for $^{114}$Sn, and three for $^{112}$Sn. Furthermore, 5, 13 and 12 scans of the odd-even $^{115,117,119}$Sn isotopes were performed, respectively. The IS of the FES for the full isotopic chain and HFS constants of $^{115,117,119}$Sn has been extracted. Our results are compared with literature results \cite{gorges2019,yordanov2020,Vazquez2018,anselment1986} and are presented in Table \ref{tab:Sn I RIS IS 1st}. The results are in agreement, albeit the $^{114}$Sn case has a slightly larger deviation than 2$\sigma$ with respect to the results of \cite{anselment1986}.

For the  SES probing, a total of 19 scans were recorded. Here 15 scans include the $^{114,116,118,120,122,124}$Sn even-even isotopes and 14 scans include the lowest abundance $^{112}$Sn even-even isotope. Our IS results from these measurements are presented in Table \ref{tab:Sn I RIS IS and HFS 2nd}.

The reliability of the results for these novel SES probing measurements was once more tested via King plot analysis with reference IS measurements using the $5p^2$ $^1S_0 \rightarrow$ $5p6s$ $^1P^o_1$ atomic transition with reported $F_{453}$ = 2.790(23)$\,$GHz/fm$^2$ and $M_{453}$ = 
-724(21) GHz*u factors in \cite{Vazquez2018}. The resulting King plot with our data is presented in the Figure \ref{fig:King plot Sn}.

From the linear fit of the data, the slope $\frac{F_{812}}{F_{453}}$ and intercept $M_{812} - \frac{F_{812}}{F_{453}}M_{453}$ in Figure \ref{fig:King plot Sn} allows extracting the $F_{812}$ and $M_{812}$ factors of the 812 nm transition. The extracted field-shift ratio from the linear fit slope is $F_{812}$/$F_{453}$ = -0.363(138) and the intercept corresponds to 107(223) GHz*u. The resulting 812 nm transition $F_{812}$ and $M_{812}$ factors along with their uncertainties have been determined to be 

\begin{align*}
    F_{812} = -1012(394) \: \mathrm{MHz\ fm^{-2}} \\
    M_{812} = 369(340) \: \mathrm{GHz*u}. \\
\end{align*}

\end{multicols}

\begin{table}[h!]
    \footnotesize
    \centering
    \begin{tabular}{c c c c c c c c c}
        \hline
        \hline
        \multicolumn{2}{c}{} & \multicolumn{3}{c}{Isotope shifts (MHz) $\Delta\nu_{\mbox{WA}}^{A{\prime},124}$} & &\multicolumn{3}{c}{HFS $A$ coefficients of the FES (MHz)}  \\
        \multicolumn{9}{c}{} \\
         & & \multicolumn{3}{c}{$5s^25p^2$ $^3P_0$ $\rightarrow 5s^25p6s$ $^3P_1$} & & \multicolumn{3}{c}{$5s^25p6s$ $^3P_1$} \\
        \hline
        A* (amu) & $I^{\pi}$ & This work & \cite{gorges2019,Vazquez2018} & \cite{anselment1986} & & This work & \cite{yordanov2020} & \cite{anselment1986}\\
        \cline{3-5} 
        \cline{6-9} 
        122  & 0 & -233(62)  & -206.3(75) [10] & -205.80(21) & & & & \\
        120  & 0 & -429(14)  & -448.6(80) [22] & -441.15(15) & & & & \\
        119  & \: 1/2$^+$ & -622(37)  & -619.7(28) [28] & -620.74(19) & & -5041(24) & -5011(3) & -5007.95(17)\\
        118  & 0 & -689(23)  & -695.0(67) [33] & -711.39(21) & & & &\\
        117  & \: 1/2$^+$ & -921(27)  & -906.6(50) [40] & -912.58(19) & & -4816(30) & -4783(2) & -4785.45(17)\\
        116  & 0 & -1002(32) & -1007.6(79) [45] & -1017.19(21) & & & & \\
        115  & \: 1/2$^+$ & -1173(76) & -1250.4(28) [52] & -1246.07(19) & & -4331(169) & -4394(2) & -4394.16(14)\\
        114  & 0 & -1307(16) & -1335.6(62) [58] & -1341.83(21) & & & & \\
        112  & 0 & -1647(18) & -1652.1(56) [71] & -1659.44(21) & & & & \\
        \hline
        \hline
    \end{tabular}
    \caption{(Left) Extracted weighted average IS values $\Delta\nu_{WA}^{A^{\prime},124}$ = $\nu_{\mbox{WA}}^{A^{\prime}}$ - $\nu_{\mbox{WA}}^{124}$ of stable $^{112-122}$Sn isotopes with respect to $^{124}$Sn for the FES $5s^25p^2$ $^3P_0$ $\rightarrow 5s^25p6s$ $^3P_1$ (286.4 nm) atomic transition from this work and literature \cite{gorges2019,Vazquez2018,anselment1986}. (Right) Hyperfine structure constants $A$ for the FES of $^{115,117,119}$Sn from this work and literature \cite{yordanov2020,anselment1986}. The $\sigma(\Delta\nu_{\mbox{WA}}^{A{\prime},124})$ from this work indicated in parenthesis represents the combined statistical and systematic uncertainties, for \cite{gorges2019,Vazquez2018} the parenthesis and square brackets indicate statistic and systematic uncertainties and for \cite{anselment1986} the parenthesis indicate the standard deviation of the data set. The $\sigma(A)$ indicated by parenthesis in all cases represents the standard deviation of the data set.}
    \label{tab:Sn I RIS IS 1st}
\end{table}

\begin{table}[h!]
    \footnotesize
    \centering
    \begin{tabular}{c c c}
        \hline
        \hline
        \multicolumn{3}{c }{Isotope shifts (MHz) $\Delta f_{\mbox{WA}}^{A{\prime},122}$} \\
        \multicolumn{3}{c}{} \\
        & & \multicolumn{1}{c }{$5p6s$ $^3P^o_1 \rightarrow 5p6p$ $^3P$} \\
        \hline
        A* (amu) & $I^{\pi}$ & This work \\
        \cline{3-3} 
        124 & 0 & -58(13) \\
        120 & 0 & 46(10) \\
        118 & 0 & 116(15) \\
        116 & 0 & 206(14)\\
        114 & 0 & 281(13)\\
        112 & 0 & 366(19)\\
        \hline
        \hline
    \end{tabular}
    \caption{Extracted weighted average IS values $\Delta\nu_{WA}^{A^{\prime},122}$ = $\nu_{\mbox{WA}}^{A^{\prime}}$ - $\nu_{\mbox{WA}}^{122}$ of the stable even-even $^{112-124}$Sn isotopes with respect to $^{122}$Sn from the SES $5s^25p6s$ $^3P_1$ $\rightarrow 5s^25p6p$ $^3P$ (811.6 nm) atomic transition. The $\sigma(\Delta\nu_{\mbox{WA}}^{A{\prime},122})$ indicated in parenthesis represents the combined statistical and systematic uncertainties.}
    \label{tab:Sn I RIS IS and HFS 2nd}
\end{table}

\begin{Figure}
    \centering
    \includegraphics[trim=0.1cm 0.1cm 0.1cm 0.1cm,clip,width=0.49\linewidth]{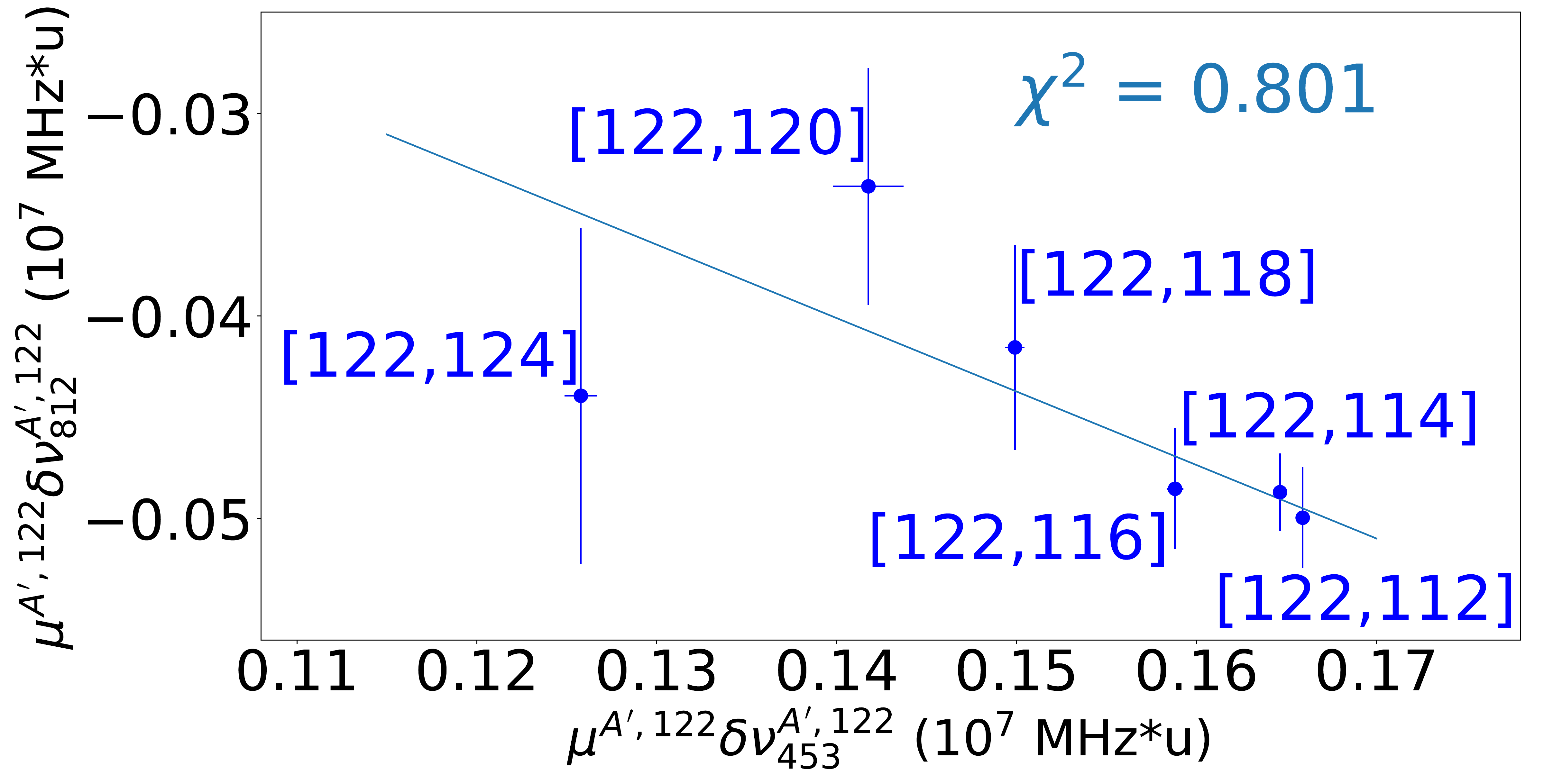}
    \captionof{figure}{King plot of modified IS of 812 nm transition as a function of modified IS of 453 nm transition from \cite{Vazquez2018}.}
    \label{fig:King plot Sn}
\end{Figure}

\begin{multicols}{2}

\section{Conclusions}
\label{ch:concl & outl}

The recent development work in the GISELE Ti:Sa laser laboratory in the framework of the S$^3$-LEB project has resulted in successful RIS measurements of the stable erbium and tin isotopic chains. These measurements quantify the previously reported qualitative performance of the SM injection-locked Ti:sa system \cite{romans2022} under experimental conditions. In detail, the IS as well as the HFS constants of the $4f^{12}6s^2$ $^3H_6 \rightarrow 4f^{12}(^3 H)6s6p$ $J = 5$ (415 nm) atomic transition in erbium have been extracted. The obtained $^{167}$Er ground and excited state HFS constants agree with those from the literature within a 1$\sigma$ uncertainty interval. A King plot analysis has been used to test the reliability of the measurements from this campaign using the reported IS measurement for the $4f^{12}6s^2$ $^3H_6 \rightarrow$ $4f^{12}$($^3H_6$)$6s6p$ $^3P_1$ (583 nm) atomic transition. Furthermore, the second King plot analysis using muonic X-ray measurements was used to obtain $F$ and $M$ atomic parameters for the 415 nm transition.

From the obtained spectra of stable tin isotopes the obtained IS as well as the HFS constants for the $5s^25p^2$ $^3P_0$ $\rightarrow 5s^25p6s$ $^3P_1$ (286.4 nm) atomic transition have been compared with the reported literature values. The IS results from this work are in an agreement with the literature values apart from the $^{114}$Sn case, most likely due to the low statistics. The HFS constants from this work agree with the literature values within a 2$\sigma$ interval. In addition, new measurements for the IS in the $5s^25p6s$ $^3P_1$ $\rightarrow 5s^25p6p$ $^3P$ (812 nm) atomic transition in even-even tin isotopes have been performed. The atomic parameters $F$ and $M$ have been extracted via a King plot analysis for this work reliability test using the reference IS measurement of the $5p^2$ $^1S_0 \rightarrow$ $5p6s$ $^1P^o_1$ (453 nm) atomic transition. For further laser spectroscopy of the SES an introduction of dithering of the FES broad-band laser resonator would allow to $smear$ out the laser modes, which will be implemented in the future at the GISELE laboratory. In the case of long-lived excited states also a delay between the successive RIS steps can improve the resolution without a loss in efficiency \cite{groote2017}. 

The GISELE Ti:sa laser laboratory has been developed for high resolution laser spectroscopy studies and the obtained results on erbium and tin of IS and HFS constants have been found in agreement with the literature. These measurements indicate that the laser system is capable of the intended first physics cases using the in-gas-jet laser ionization
spectroscopy method at the SPIRAL2-GANIL S$^3$ facility, and that the systematic uncertainties are under control. 

As a technical outlook the development of a SM Ti:sa system seed laser for improved continuous scanning range is essential. The achieved short mode-hop-free scanning ranges of 1 - 3 GHz obtained with current ECDL seed lasers make experiments difficult even in offline conditions. Furthermore, to enable simultaneous laser spectroscopy studies with high resolution at both the GISELE laboratory and the S$^3$-LEB, another SM injection-locked Ti:sa laser system is currently commissioned. At last, the continuous implementation and tests of upgrades of the existing Ti:sa systems at GISELE laboratory are vital to obtain highest laser powers, narrowest emission spectrum in fundamental and higher harmonics as well as a long term stable running conditions. In addition, more efficient ways of producing higher harmonic light, e.g. via using periodically poled nonlinear crystals, can be investigated.      

\section{Acknowledgements}

This setup results from the collaborative work of the IGLIS newtork, grouping many research centers and universities including CEA-Saclay (IRFU), CERN (CRIS), GANIL, GSI, IBS-RISP, IJCLab , IMP, JAEA, Johannes Gutenberg-Universität Mainz (Institut für Physik/LARISSA), JINR (GALS), JYFL (IGISOL/MARA), KEK (KISS), KU Leuven, MSU, Nagoya University, Normandie Université (LPC Caen), Peking University, RIKEN (SLOWRI/PALIS), TRIUMF (TRILIS), Université de Strasbourg (IPHC), University of Manchester, University of Tsukuba and Laboratoire de Physique des 2 infinis Irène Joliot-Curie (IJCLab) (for more details about IGLIS collaboration please refer to our network page \cite{IGLISweb}). 

S$^{3}$ has been funded by the French Research Ministry, National Research Agency (ANR), through the EQUIPEX (EQUIPment of EXcellence) reference ANR-10EQPX- 46, the FEDER (Fonds Europe\'en de De\'veloppement Economique et Re\'gional), the CPER (Contrat Plan Etat Re\'gion), and supported by the U.S. Department of Energy, Office of Nuclear Physics, under contract No. DE-AC02-06CH11357 and by the E.C.FP7-INFRASTRUCTURES 2007, SPIRAL2 Preparatory Phase, Grant agreement No.: 212692.
S$^{3}$-LEB: This project has received funding from the French Research Ministry through the ANR-13-B505-0013, the Research Foundation---Flanders (FWO)---under the International Research Infrastructure program (nr. I002219N), the Research Coordination Office---KU Leuven---the European Research Council (ERC-2011-AdG-291561-HELIOS) and the European Union’s Horizon 2020 research and innovation program under grant agreement No 654002. This project has received funding from the European Union’s Horizon 2020 research and innovation programme under grant agreement No 861198–LISA–H2020-MSCA-ITN-2019

\printbibliography[heading=bibnumbered, title={References}]

\end{multicols}
\end{document}